\documentclass[sigconf,anonymous=False,review=False,natbib=True]{acmart}
\renewcommand\footnotetextcopyrightpermission[1]{} 
\usepackage{tabularx, booktabs}
\usepackage{graphicx}
\usepackage{multirow}
\usepackage{ulem}
\usepackage{setspace}
\usepackage{xcolor}
\usepackage{array}
\usepackage{balance}
\usepackage{enumitem}
\usepackage{booktabs}
\usepackage{ulem}
\usepackage{CJKutf8}
\usepackage{makecell}
\usepackage{geometry}
\usepackage{subcaption}
\usepackage{acronym}

\acrodef{IR}{Information Retrieval}
\acrodef{MLP}{multilayer perceptron}
\acrodef{SERP}{Search Engine Result Page}
\acrodef{ERP}{event related potential}
\acrodef{EEG}{electroencephalogram}
\acrodef{DT}{Gradient Boosting Decision Tree}
\acrodef{SST}{SST-EmotionNet}
\acrodef{AUC}{Area Under Curve}
\acrodef{IN}{Information Need}
\acrodef{fMRI}{functional magnetic resonance imaging}
\acrodef{BCI}{brain–computer interface}
\acrodef{BP}{band power}
\acrodef{DE}{differential entropy}
\acrodef{RASM}{rational asymmetry}
\acrodef{DASM}{differential asymmetry}
\acrodef{SVM}{support vector machines}
\acrodef{DBN}{deep belief networks}
\acrodef{KNN}{k-Nearest Neighbors}
\acrodef{HBG}{Height-Biased Gain}

\AtBeginDocument{%
  \providecommand\BibTeX{{%
    \normalfont B\kern-0.5em{\scshape i\kern-0.25em b}\kern-0.8em\TeX}}}

\makeatletter
\def\hlinew#1{%
  \noalign{\ifnum0=`}\fi\hrule \@height #1 \futurelet
   \reserved@a\@xhline}
\makeatother

\setcopyright{acmcopyright}
\copyrightyear{2021}
\acmYear{2021}
\acmDOI{10.1145/1122445.1122456}

\acmConference[WSDM '15]{WSDM '15: The 15th ACM International Conference on Web Search Search and Data Mining}{February 21--25, 2022}{Phoenix, Arizona}

\begin{document}
\fancyhead{}

\title{Why Don't You Click: Neural Correlates of Non-Click Behaviors in Web Search}

\author{Ziyi Ye}
\email{yeziyi1998@gmail.com}
\affiliation{%
  \institution{BNRist, DCST, Tsinghua University}
  \city{Beijing}
  \country{China}}

\author{Xiaohui Xie}
\email{xiexh_thu@163.com}
\affiliation{%
  \institution{BNRist, DCST, Tsinghua University}
  \city{Beijing}
  \country{China}}

\author{Yiqun Liu}
\email{yiqunliu@tsinghua.edu.cn}
\affiliation{%
  \institution{BNRist, DCST, Tsinghua University}
  \city{Beijing}
  \country{China}}

 \author{Xuancheng Li}
\email{lixuanch18@mails.tsinghua.edu.cn}
\affiliation{%
  \institution{BNRist, DCST, Tsinghua University}
  \city{Beijing}
  \country{China}}

\author{Jiaji Li}
\email{jiajili@link.cuhk.edu.cn}
\affiliation{%
  \institution{SDC, The Chinese University of Hong Kong, Shenzhen}
  \city{Beijing}
  \country{China}}
  
  \author{Zhihong Wang}
\email{wangzhh629@mail.tsinghua.edu.cn}
\affiliation{%
  \institution{BNRist, DCST, Tsinghua University}
  \city{Beijing}
  \country{China}}
 
 \author{Xuesong Chen}
\email{chenxuesong1128@163.com}
\affiliation{%
  \institution{BNRist, DCST, Tsinghua University}
  \city{Beijing}
  \country{China}}

 \author{Min Zhang}
\email{	z-m@tsinghua.edu.cn}
\affiliation{%
  \institution{BNRist, DCST, Tsinghua University}
  \city{Beijing}
  \country{China}}

 \author{Shaoping Ma}
\email{	msp@tsinghua.edu.cn}
\affiliation{%
  \institution{BNRist, DCST, Tsinghua University}
  \city{Beijing}
  \country{China}}

\renewcommand{\shortauthors}{Ye, et al.}

\begin{CJK}{UTF8}{gbsn}
\begin{abstract}

Web search heavily relies on click-through behavior as an essential feedback signal for performance improvement and evaluation. 
Traditionally, click is usually treated as a positive implicit feedback signal of relevance or usefulness, while non-click~(especially non-click after examination) is regarded as a signal of irrelevance or uselessness. 
However, there are many cases where users do not click on any search results but still satisfy their information need with the contents of the results shown on the Search Engine Result Page~(SERP). 
This raises the problem of measuring result usefulness and modeling user satisfaction in ``Zero-click'' search scenarios.

Previous works have solved this issue by~(1) detecting user satisfaction for abandoned SERP with context information and~(2) considering result-level click necessity with external assessors' annotations. 
However, few works have investigated the reason behind non-click behavior and estimated the usefulness of non-click results.
A challenge for this research question is how to collect valuable feedback for non-click results.
With neuroimaging technologies, we design a lab-based user study and reveal differences in brain signals while examining non-click search results with different usefulness levels. 
The findings in significant brain regions and electroencephalogram~(EEG) spectrum also suggest that the process of usefulness judgment might involve similar cognitive functions of relevance perception and satisfaction decoding.
Inspired by these findings, we conduct supervised learning tasks to estimate the usefulness of non-click results with brain signals and conventional information~(i.e., content and context factors). 
Results show that it is feasible to utilize brain signals to improve usefulness estimation performance and enhancing human-computer interactions in ``Zero-click'' search scenarios.
  
\end{abstract}
\end{CJK}
\settopmatter{printacmref=false}




\maketitle
\begin{CJK}{UTF8}{gbsn}
\section{introduction}
\begin{figure}[t]
  \centering
  \includegraphics[width=1\linewidth]{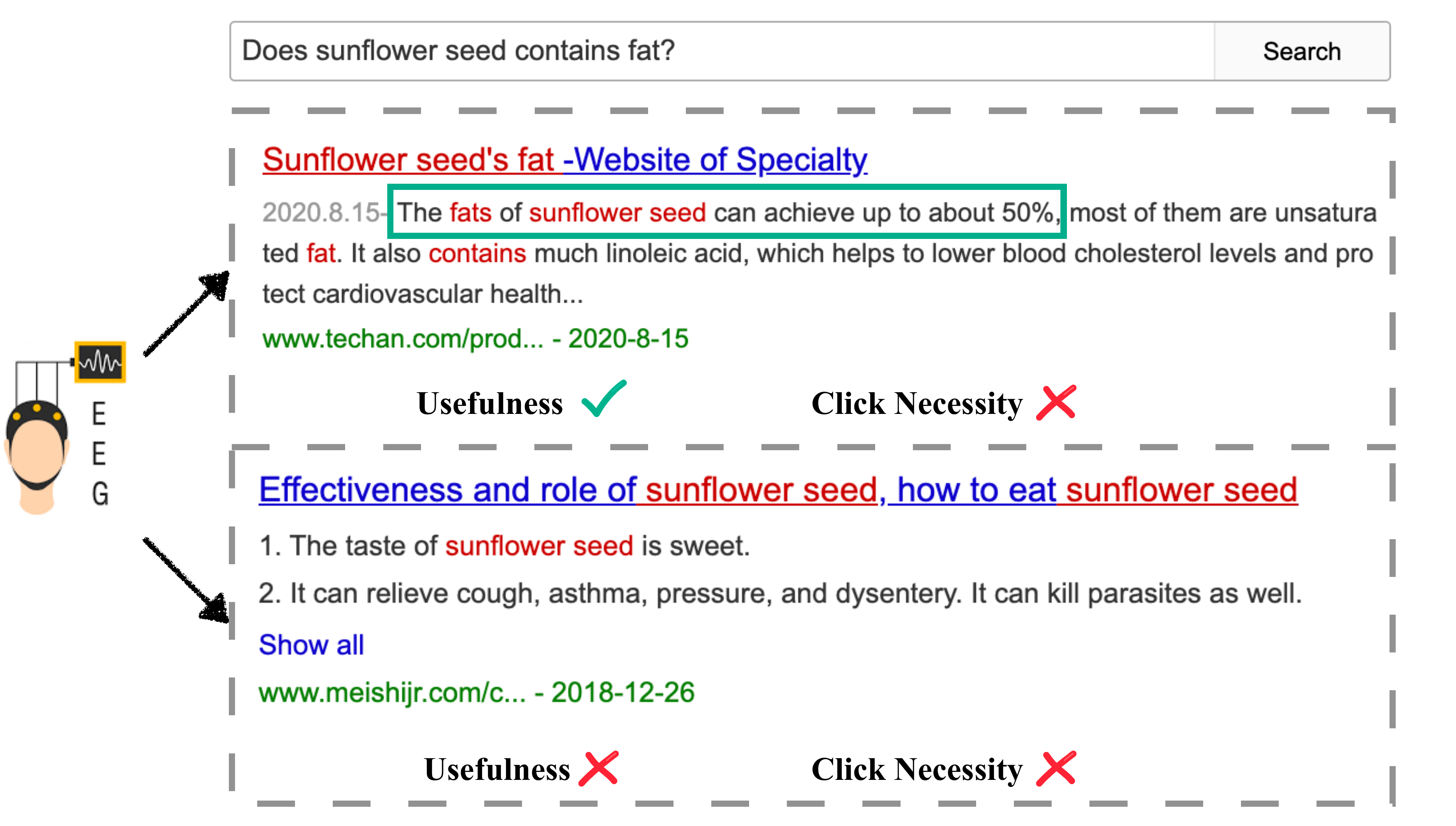}
  \caption{Examples of non-click search results. The first result is helpful to satisfy the user's information need with its snippets and unnecessary to click.} 
  \label{fig:non-click}
  \Description[]{}
  \vspace{-3mm}
\end{figure}

The \ac{IR} community has a long tradition of using click-through behavior as vital user feedback for search evaluation~\cite{joachims2003evaluating} and relevance modeling~\cite{radlinski2005query,joachims2017accurately}.
In these works, click is usually considered as a positive signal , while a non-click result~(especially non-click after examination) is usually regarded as irrelevant or useless.
However, commercial search engines have attempted to add features that can satisfy users’ \ac{IN} directly through the \ac{SERP}.
This kind of information acquisition process is called ``Zero-click'' search~\footnote{https://www.searchmetrics.com/glossary/zero-click-searches/} and a phenomenon called ``good abandonment'' would happen when a user satisfies his \ac{IN} without clicks on any results. 
Figure~\ref{fig:non-click} gives examples of two non-click search results: both results are unnecessary to click, and the first result is helpful to satisfy the user's \ac{IN} with its snippets.  

Recent studies have shown that ``Zero-click'' Google searches rose to nearly 65\% in 2020, up from 50.33\% in 2019~\footnote{https://sparktoro.com/blog/in-2020-two-thirds-of-google-searches-ended-without-a-click/}.
This phenomenon occurs more frequently for portable devices~\cite{li2009good} and proactive search systems that can extract better snippets or special components for a better information access experience~\cite{sakai2011click}. 
Therefore, understanding why the user does not click a search result becomes a vital challenge and attracts much attention.
On the one hand, existing efforts focus on detecting page-level ``good abandonment''  with search logs~\cite{song2014context}, \ac{SERP} content~\cite{williams2016detecting}, and user interactions~\cite{diriye2012leaving,williams2016detecting}. 
On the other hand, ~\citet{luo2017evaluating} propose the concept of ``click necessity'', which is a golden standard acquired by external assessors' annotations.
It is later adopted to model non-click behaviors in search evaluation~\cite{luo2017evaluating} and click model construction~\cite{mao2018constructing}.

As mentioned above, good abandonment prediction is a page-level task, and click necessity does not directly reflect whether a search result is good or bad.
However, few studies have delved into understanding the result-level usefulness of non-click ones.
The main challenge is that conventional user interactions like cursor and eye-movements can hardly uncover the attention status and cognitive process that happened in the human brain straightforwardly.
And they might not be effective enough: a previous experiment shows that classification based on cursor behavior and dwell time achieves binary prediction performance~($F_{0.5}$) of 0.505 in good abandonment detection~\cite{diriye2012leaving}.
Nevertheless, \ac{SERP} content information achieves that of 0.617. 
Thus there is still room for improving the performance with user feedback.

Recently, the rapid development of neuroimaging technology~(e.g., \ac{EEG} and \ac{fMRI}) makes it feasible to explore brain activities during the search process.
Extensive studies have applied neurological devices to explore the nature of relevance judgment~\cite{allegretti2015relevance, gwizdka2017temporal} and realization of \ac{IN}~\cite{moshfeghi2019towards}.
These studies are designed to simulate the examination procedures in the landing page and usually leave out the interactions on \ac{SERP}.
However, along with the growth of the ``Zero-click'' scenarios, understanding the interactions before a click, especially the reason behind non-click behavior, is a vital challenge.

In this paper, we delve into the cognitive processes of non-click behavior and explore the effectiveness of different information sources: brain signals and conventional information~(i.e., content and context factors in this paper) for usefulness estimation:
\begin{itemize}
	\item \textbf{RQ1}: Are there any detectable differences in brain activities while examining non-click search results with different usefulness? If yes, 
	\item \textbf{RQ2}: To what extent can we estimate the usefulness of non-click results with additional information sources of brain signals? And, 
	\item \textbf{RQ3}: How do different information sources and experimental settings contribute to the performance of usefulness estimation?
\end{itemize}
To shed light on these research questions, we conduct a lab-based user study to investigate non-click behavior.  
Participants are required to perform pre-defined search tasks while an \ac{EEG} device is applied to collect brain activities during this process.
With analysis of the \ac{EEG} spectrum, we find brain activities vary with non-click behaviors regarding the results' usefulness. 
Notably, we find significant correlations between \ac{EEG} band power and result usefulness, especially in brain regions of left temporal, frontal, and occipital.
These findings indicate that usefulness judgment involves several cognitive functions that have something in common with relevance perception~\cite{moshfeghi2013understanding} and satisfaction decoding~\cite{moshfeghi2019towards}.
Additionally, these findings also illustrate the possibility of utilizing brain signals for the usefulness estimation of non-click results.

To verify the effectiveness of brain signals in the usefulness estimation task, we conduct extensive experiments based on brain signals with different settings: user-independent and task-independent.
Experimental results demonstrate that brain signals can bring a significant additional improvement of 6.8\%~(user-independent) and 13.5\%~(task-independent) in terms of \ac{AUC} compared to previous usefulness estimation models based on content and context information~\cite{mao2017understanding}.
These experimental results also illustrate that brain signals are valuable feedback during \ac{SERP} examination and demonstrate the possibility of constructing a more proactive ``Zero-click'' search system with real-time \ac{BCI}.
%

\section{related work}

\subsection{Zero-click Search}
``Zero-click'' refers to the situation that the \ac{SERP} successfully and entirely satisfies the \ac{IN}, without the necessary to click on a search result.
Recently, commercial search engines have been attempting to improve user experience by extracting high-quality snippets or creating enhanced search results so that a user can pay as little effort~(including click) as possible to access the \ac{IN}.  
Therefore, ``Zero-click'' search plays an important role in real-world \ac{IR} and attracts much attention.

To understand the non-click behaviors in Web search, recent researches have concentrated on ``good abandonment'', which is when the user’s \ac{IN} is successfully addressed with no need to click on a result or refine the query.
For instance, \citet{li2009good} approximate the prevalence of good abandonment in desktop and mobile search logs and find that a large amount of abandonment behavior is good abandonment, especially in mobile search.
Additionally, some researchers detect and predict good abandonment in desktop~\cite{diriye2012leaving} and mobile~\cite{williams2016detecting,wu2020credibility}, with the help of page content and user interaction with \ac{SERP}.
As for search evaluation, \citet{khabsa2016learning} propose an online metric that accounts for good abandonment to overcome the shortcomings of simply utilizing click as positive signals.
Going one step further, researchers explore an ideal situation in the future, which is called ``Zero Query''.
Instead of waiting for the user to enter a query and click search results, the system decides when and what to provide~\cite{sakai2012towards}.
The first step towards building a feasible and practical search engine as they designed is to predict result usefulness with more effective user feedback.

Another trend of researches address this issue by considering result-level click necessity, which usually needs external assessors' annotations.
For instance, \citet{luo2017evaluating} propose a concept of click necessity and a novel metric called \ac{HBG} for mobile search evaluation.
\citet{zhang2018relevance} propose a joint relevance estimation model that using click necessity as features and achieving better performance than state-of-the-art ranking solutions.
In addition to collecting external assessors' annotations of click necessity, some researches consider click necessity as a trainable value to construct click models which achieve better performance, especially in mobile search~\cite{mao2018constructing}.
However, click necessity doesn't resolve the usefulness estimation for non-click results directly.
What it can account for is that when a search result is click unnecessary, it is either useless or contains direct answers in the snippet.  

Our contributions in this paper complement existing work on ``Zero-click'' search by uncovering the brain activities of non-click behaviors during usefulness judgment. 
Then we leverage brain signals as user feedback and demonstrate its effectiveness for estimating result usefulness.

\subsection{Usefulness of Search Result}
Usefulness is a key concept in the user-centric evaluation of Web search.
In contrast to relevance, which is often annotated by external assessors, usefulness represents users’ opinions about whether search results can meet their \acp{IN}~\cite{voorhees2001philosophy}.
\citet{mao2016does} find that there exist many cases that high relevance may not necessarily mean the document is useful for the user.
And they reveal that usefulness has a higher correlation with user satisfaction than relevance annotated by external assessors.
With such findings, they further propose models for usefulness judgment prediction in desktop search scenarios~\cite{mao2017understanding} and mobile search scenarios~\cite{mao2018investigating}.

However, few studies have delved into understanding the usefulness judgment for non-click results.
One of the challenges is the interactions in the landing page, which contain valuable feedback such as dwell time and mouse movement, are absent for non-click results. 
What we add on top of these works is that we collect brain signals during the examination of non-click results to uncover this problem.
In addition, to verify the effectiveness of collected brain signals, we evaluate and compare between the usefulness estimation model based on brain signals and conventional features propose by \citet{mao2017understanding}, which is described in detail in Section~\ref{Features}.

\subsection{Neuroscience \& IR}
There is an increasing number of interdisciplinary research literature focusing on the intersection of neuroscience and \ac{IR}.
The contributions of these works are two folds: (1)~revealing the neurological process in \ac{IR} and (2)~leveraging brain signals as user feedback for evaluation.  

For the first contribution, several studies investigate how the different cognitive components of \ac{IR} emerge from activity in the brain.
For example, \citet{moshfeghi2013understanding,moshfeghi2016understanding,moshfeghi2019towards} and \citet{pinkosova2020cortical} conduct a series of studies using brain signals to unravel the nature of a set of core notions, such as relevance and \ac{IN}.
They demonstrate the distributed network of brain regions associated with these concepts and related IR tasks.
Insightful findings are obtained such as (1)~ \ac{IN} reflects a switch in the human brain to acquire external information sources and (2)~ relevance is a graded phenomenon.

For the second contribution, recent years have witnessed the field of \ac{IR} being increasingly more interested in utilizing brain signals as user feedback for better human computer interaction.
For instance, \citet{gwizdka2017temporal} conduct extensive studies to infer page relevance using \ac{EEG} or in combination with eye-movements.
Their classification results that models using \ac{EEG} features can achieve an improvement of 20\% in \ac{AUC} compared against that of an untrained model.
\citet{kim2019erp} extend their work into the field of visual shots and classify topical relevance with \ac{EEG} algorithm.
However, their experimental design is different from a real search scenario where users would examine search results before clicking results and reading the landing page. 

What we add on top of these works is that we conduct a user study where participants can freely interact with search results.
On this basis, we delve into understanding the ``Zero click'' scenario and detecting the usefulness of non-results with brain signals.

\section{Data collection}
In this section, we introduce the design of our user study and the collected dataset~\footnote{The data and code will be publicly available after the review process.}.

\subsection{User study tasks}
We first select 150 queries from the SRR~(Search Result Relevance) dataset~\cite{zhang2018relevance} for our user study.
We use this dataset for two main reasons: (1)~It contains a large number of real-life query logs, screenshots of search results, and landing pages. Each query has ten corresponding results. (2)~It provides human annotations of result type according to presentation styles.
To ensure that a query has a greater probability to cause ``Zero-click'' scenario and is understandable, the sampling process is based on several criteria: 
(1)~The query should have no click interaction in the log.
(2)~A sampled query should have a clear and unambiguous description.

After that, we recruit 15 assessors to annotate the search results of selected queries.
For each result, the click necessity~(binary) and the usefulness~(five-point Likert scale) are judged by at least three different assessors, and their mean click necessity are applied for later selection. 
Finally, 90 queries remain since they contain at least five results with lower annotated click necessity~($\le$ 0.5).
We reserve these tasks with the expectation of collecting more non-click behaviors in the user study.
For each task, we generate a task description manually according to the query and collect the corresponding search results in the dataset for our study.

\subsection{Participants}
We recruit 18 college students aged from 19 to 26~(M~\footnote{Mean value.} = 21.56, SD~\footnote{Standard deviation.} = 1.82).
The number of participants are analogous to previous \ac{EEG}-based user studies~(e.g., 15 participants in~\cite{duan2013differential} and 20 participants in ~\cite{allegretti2015relevance}).
There are ten males and eight females, who mainly major in computer science, physics, arts, and engineering. 
All the participants are familiar with the basic usage of search engines, and all of them report using search engines daily or once in two days.
The whole task takes about two hours to complete: 50 minutes for preparation and rest, 60 minutes for the main task, and 10 minutes for the questionnaire procedure. 
And each participant would gain \$30 after they complete all the tasks seriously.

\subsection{Procedure}

\begin{figure}[t]
  \centering
  \includegraphics[width=1\linewidth]{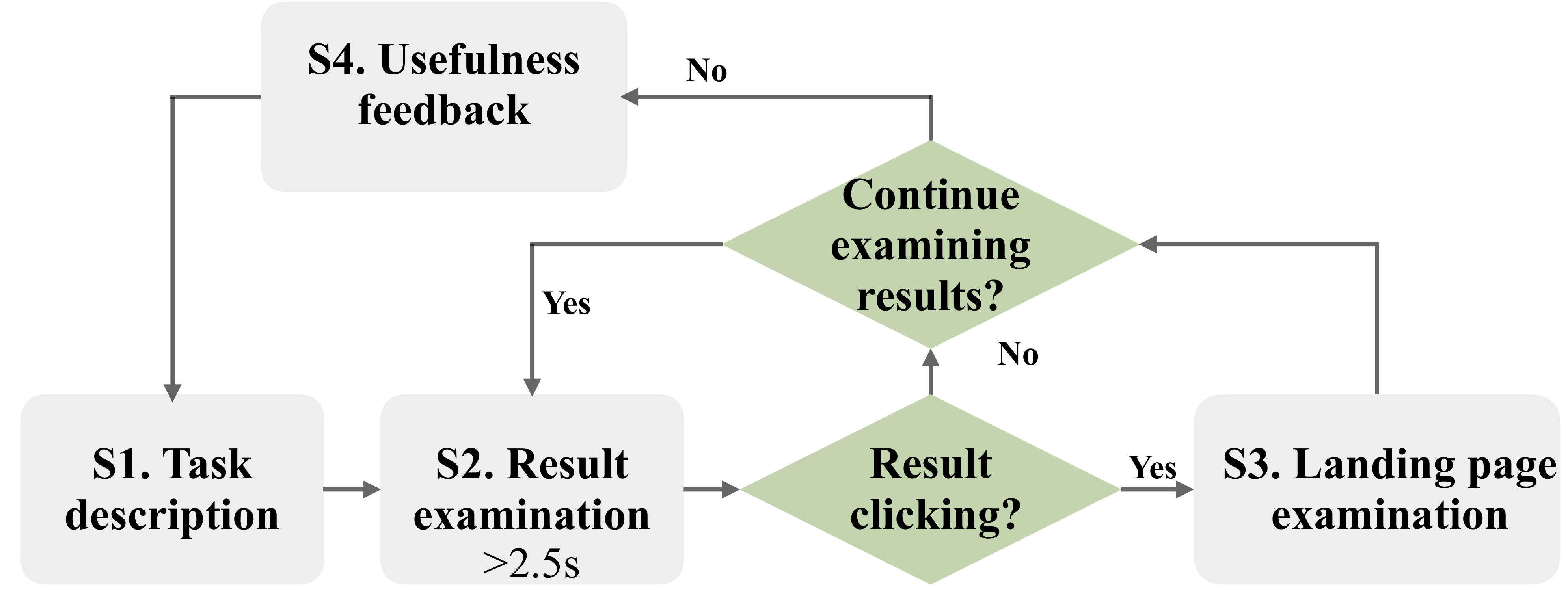}
  \caption{The procedure of a search task. If participants choose not to click in S2, the result is a non-click one.} 
  \label{fig:user study}
  \Description[]{}
\end{figure}

This user study adheres to the ethical procedures for the protection of human participants in research and is approved by \textbf{anonymized}.
In the beginning, participants fill in an entry questionnaire to report demographic information and sign an informed consent about security and privacy protection. 
Then they read user study instructions about the procedure of each search task during the user study. 
Before entering the main step, participants undergo a training step with two tasks to ensure they are familiar with the procedure.
Each participant is instructed to complete a search task using a web browser in our experimental system developed with Django for a task in the training step and the main step.
And the participants are supposed to seriously accomplish the search tasks as many as possible in 60 minutes. 
They are allowed to rest between tasks while the rest time would not be included in the time limit.  

Figure~\ref{fig:user study} illustrates the procedure of each search task in the main step.
The trials follow the same order of steps, i.e., S1 to S4 shown in Figure~\ref{fig:user study}: 

(S1)~Participants view a task description randomly selected from the dataset. 
Once they fully understand the question, they can press a button and enter the second step. 

(S2)~A fixation cross is presented for 1.5 seconds on the screen center to capture participants' attention and indicate the location of the forthcoming result.
Then a randomly selected search result screenshot~(Figure~\ref{fig:non-click} gives two examples) is displayed, lasting for 2.5 seconds.
After that, three response choices, i.e., ``skip'', ``click'', and ``end search'', are presented on the screen while the search result is still shown. 
This procedure, following the previous works~\cite{moshfeghi2016understanding,2108.01360}, ensures that brain activity related to the motor response of moving the cursor and click the button would not be contained during the 2.5 seconds.
And we would use brain signals recorded in this time interval for further analysis and experiments.

(S3)~If participants choose ``click'' in~(S2), the landing page of the corresponding result will be presented. 
After examining the lading page, the participant can either end the search or continue to examine the next result in this step.

(S4)~If the participant decides to end the search in~(S2) or~(S3), they are presented with an end-mark page.
On this page, they are required to give a brief answer to the search task via voice input and report their perceived difficulty~(five-point Likert scale) to the search task and usefulness feedback~(four-point Likert scale) to each result.

We randomize the tasks' order for each participant and display the screenshot of each result in a randomized sequence.
A pilot study, which involved four additional users, is conducted ahead to adjust the settings including the display time of fixation cross and result, amount of training trails, etc.
Note that in our experimental paradigm, the search result is displayed one by one.
We apply this paradigm to collect brain signals and behavior responses for each specific result and left the investigation on the whole page as future work.  

\subsection{Apparatus}
Our study uses a desktop computer that has a 27-inch monitor with a resolution of 2,560×1440 and Google Chrome browser.
A Scan NuAmps Express system~(Compumedics Ltd., VIC, Australia) and a 64-channel Quik-Cap~(Compumedical NeuroScan) are deployed to capture the participants’ \ac{EEG} data. 
All the \ac{EEG} electrodes are placed based on the International 10–20 system.
The impedance of the electrodes is calibrated under 10 $k\Omega$ in the preparation step, and the sampling rate is set at 1,000 Hz. 
The computations and data pre-processing are performed using the Curry V8.3~(Neuroscan, TX), a widely used commercial source localization software package.

\subsection{Statistics of the Collected Data}

\begin{table}[t]
\caption{The average number of participants' responses across usefulness levels.}
\label{tab:static}
\begin{tabular}{l|llll}
\hlinew{0.8pt}
\multirow{2}{*}{Response}         & \multicolumn{4}{c}{\textbf{Usefulness}}                   \\ 
  & 1           & 2          & 3          & 4          \\ \hline
\makecell[c]{click}     & \makecell[l]{$14.0(\pm 14)$}  & \makecell[l]{$7.4(\pm 7)$}   & \makecell[l]{$10.2(\pm 11)$} & \makecell[l]{$12.8(\pm 13)$} \\ 
\makecell[c]{non-click} & \makecell[l]{$101.2(\pm 53)$} & \makecell[l]{$30.8(\pm 21)$} & \makecell[l]{$31.5(\pm 20)$} & \makecell[l]{$52.5(\pm 16)$} \\ \hlinew{0.8pt}
\end{tabular}
\end{table}


The collected dataset consists of 1252 interactions on 90 search tasks, participants examine 3.61(SD=2.24) search results for each task on average.
Note that participants do not need to examine all result given a query.
One participant averagely accomplishes 69.56~(SD=12.23) tasks and examines 250.78~(SD=56.53) search results.
Table~\ref{tab:static} presents the participants' responses~(click and non-click) across usefulness levels ranging from 1 to 4. 
We can observe that about 85.9\% of search results are non-clicked, among which 46.8\% are ``not useful at all''~(usefulness=1), followed by ``very useful''~(usefulness=4), while fewer in ``fairly useful''~(usefulness=3) and ``somewhat useful''~(usefulness=2).

\section{analyses of brain signals}

\subsection{Preprocessing of EEG data}
\ac{EEG} data commonly contains noise sources related to power line noise, eye blinks, body movement, etc., which need to be pre-processed according to standard procedures for further analysis.
The standard procedures include: re-referencing to averaged mastoids, baseline correlation, low-pass of 50Hz and high-pass of 0.5Hz filtering, artifacts removal, and down-sampling to 500 Hz. 
For artifacts removal, a parametric noise covariance model~\cite{huizenga2002spatiotemporal} is applied to preserve the data from artifacts associated with ocular, cardiac, and muscular artifacts.
Afterward, interested epochs~(brief \ac{EEG} segment, 2,500 ms in our experimental settings) are extracted according to the display time point of each search result, and baseline correlation is applied again using the pre-stimulus period 0-1500 ms. 

\subsection{Relationship Between the \ac{EEG} Spectrum and Perceived Usefulness}

\begin{figure}[t]
  \centering
  \includegraphics[width=0.75\linewidth]{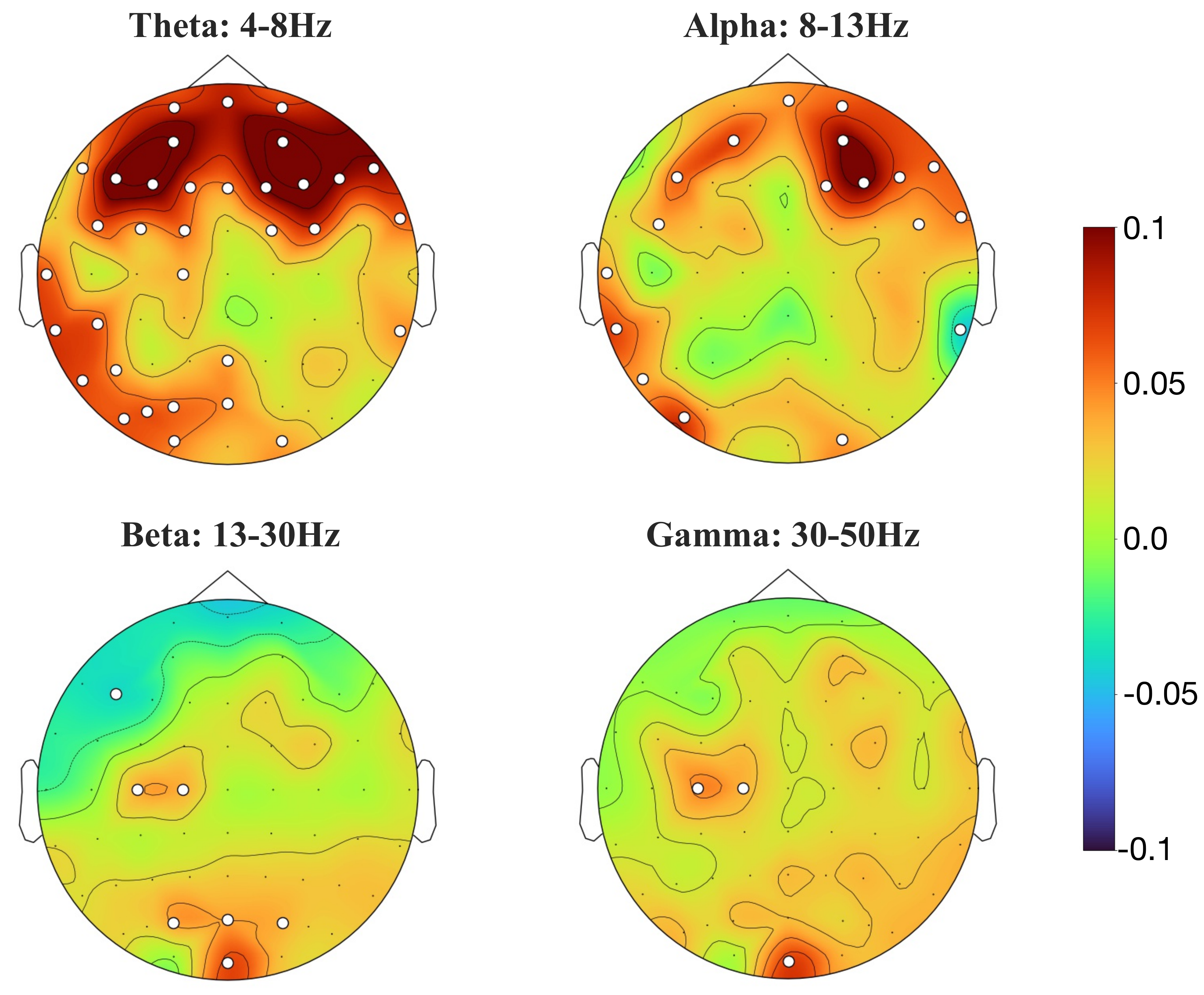}
  \caption{The correlations~(averaged over all participants) between the usefulness ratings and the spectral power. The highlighted sensors indicate that correlations are significant at the $p < 0.05$ level.} 
  \label{fig:corr topomap}
  \Description[]{}
\end{figure}

Equipped with the de-noised EEG data, we seek to answer \textbf{RQ1} by investigating the relationship of perceived usefulness and the power of the \ac{EEG} spectrum in different frequency bands.
This method is widely applied to investigate the brain activities reacting to stimuli with different types~\cite{koelstra2011deap}. 
For each channel, we extract the spectral powers of each epoch between 4 and 50 Hz according to Welch’s method and average them over the frequency bands of theta~(4-8 Hz), alpha~(8-13 Hz), beta~(13-30 Hz), and gamma~(30-50 Hz) to obtain \ac{EEG} band powers. 
The mean Pearson's correlations over all participants are measured between changes in \ac{EEG} band powers and the perceived usefulness of non-click search results, as shown in Figure~\ref{fig:corr topomap}.

From Figure~\ref{fig:corr topomap}, we can observe that more significant channels have appeared in theta and alpha bands than that of beta and gamma bands.
Theta and alpha bands are related to cognitive performance~\cite{klimesch1999eeg}.
The correlation coefficients of these channels are usually positive in those frequencies, indicating that spectral power would be significantly higher when participants examine non-click search results with higher usefulness. 
Another observation is that the correlations are high at brain regions of the frontal and left temporal in theta and alpha bands. 
Previous \ac{fMRI} studies on relevance perception~\cite{moshfeghi2013understanding} suggest that the brain activities at these regions~(i.e., frontal and left temporal), are different when processing relevant and non-relevant documents. 
These findings indicate that relevance and usefulness, though differentiated by some researchers, might be two highly connected concepts sharing similar cerebral function areas.

As for beta and gamma bands, significant correlations are found at brain regions of the occipital and left central.
Previous studies~\cite{duan2013differential,yang2020high} suggest that high-frequency \ac{EEG} signals, like beta and gamma bands, are more related to advanced cognitive functions, especially emotions.
In the scenario of search result examination, we speculate that whether the \ac{IN} is realized and whether the user is satisfied might arouse patterns of advanced cognitive functions similar to certain positive emotions.
Recent studies in \ac{IR} domain have revealed that the brain regions of medial frontal cortex, left inferior frontal gyrus, and middle occipital gyrus have significant differences in executive function depending upon whether there is a realization of \ac{IN}~\cite{moshfeghi2019towards}.
Our findings of significant brain regions in these spectral bands are partially agreed with them~(left frontal and occipital), indicating that these regions might associate with the human response to useful results that can help \ac{IN} realization.

\paragraph{Answer to \textbf{RQ1}}
To conclude, we have the following findings of brain activities: 
(1)~The powers of lower spectral bands~(theta and alpha) are related to participants' perceived usefulness, especially in frontal and left temporal.
These differences might attribute to cognitive functions similar to relevance perception.
(2)~Usefulness judgment is also correlated with advanced cognitive functions such as \ac{IN} realization and user satisfaction in high frequency \ac{EEG} spectrum~(beta and gamma).
The above analyses provide converging and insightful evidence that there are detectable differences in brain activities while examining non-click search results of different usefulness ratings.
In that regard, it is possible to improve usefulness estimation performance with collected brain signals.

\section{experimental setups}
To explore the effectiveness of brain signals in usefulness estimation, we conduct experiments to compare different models based on brain signals, content/context information, and their combinations.
Estimating the usefulness of non-click results is beneficial for better evaluation of search performance and design of a more proactive ``Zero-click'' search system. 
Since the aim of the experiment is to demonstrate the effectiveness and interpretability of \ac{EEG} signals as implicit feedback, we apply prevalent feature engineering methods and two different multichannel \ac{EEG} classification models~(i.e., \ac{DT} and \ac{SST}~\cite{jia2020sst}) in our experiment. 
We leave the investigation on other sophisticated models as future work.

\subsection{Features}
\label{Features}
This subsection elaborates selected features based on brain signals and content/context information.

\subsubsection{Brain signals}
Existing works in multichannel \ac{EEG}-based prediction extract features in the spectral domain and temporal domain~\cite{eugster2014predicting,2108.01360}.

For the spectral domain, \citet{hyvarinen1998analysis} propose \ac{DE} as \ac{EEG} features and further studies show that \ac{DE} performs better than other features including \ac{BP}, \ac{RASM}, and \ac{DASM} for \ac{EEG}-based prediction~\cite{duan2013differential} in emotion recognition tasks.
And \ac{DE} is also widely used in other multichannel \ac{EEG} classification tasks such as sleep-stage estimation~\cite{jia2020graphsleepnet}.
Therefore, we extract \ac{DE} features using Short Time Fourier Transform~(STFT) over five frequency bands~(delta: 0.5-4Hz, theta: 4-8Hz, alpha: 8-13Hz, beta: 14-30Hz, gamma: 30-50Hz) in 62 distinct channels except for two re-reference channels M1 and M2.

For the temporal domain, \citet{jia2020sst} illustrate its complementarity with commonly-used spectral features.
In the field of \ac{IR}, temporal features are shown to be associated with relevance judgments~\cite{eugster2014predicting} and decision making in information seeking~\cite{frey2013decision}.
For the above reason, we extract temporal features by down-sampling the raw \ac{EEG} data of each channel to 50Hz for our experiment.


\subsubsection{Content/Context information}
Among the researches of usefulness judgment,\citet{mao2017understanding} firstly investigate the factors that affect usefulness judgments in desktop search scenarios and extend their study into mobile devices\cite{mao2018investigating}.
They conduct usefulness estimation task using two types of factors related to: (1)~ \textit{content} information~(i.e., features of current search result) and (2)~\textit{context} information~(i.e., features of interaction history).
The features in our experiment are mostly inherited from their study, excluding those related to user behavior since the interaction with the landing page, such as dwell time, scrolling, etc., is not available for non-click results. 
In addition, we supplement the content features with \textit{result type}, which is not included in their study but is demonstrated as a related factor of click necessity by \citet{williams2016detecting}.
The feature \textit{result type} consists of 19 categories according to their presentation styles, such as ``Organic Result'' and ``Question Answering Vertical''.
Consequently, in our experiments, the \textit{content} features consist of  \textit{similarity rank of the query-result pair}~(using a BERT encoder), \textit{BM25 rank of the query-result pair}, and \textit{result type}.
Following the existing study~\cite{mao2017understanding}, we also use the \textit{context} features which contain \textit{average/max similarity rank with previous search results}, \textit{average/max/total usefulness ratings with previous search results}, and \textit{number of previous search results}.

\subsection{Models}  

Multichannel \ac{EEG} classification models can be broadly divided into two groups: topology-invariant and topology-aware.
Traditional classification models, such as \ac{SVM}, \ac{KNN}, and \ac{DT}, are belong to the group of topology-invariant type.
They do not consider the topological structure of \ac{EEG} channels when learning the representations and adopt manually designed features, especially spectral features, to circumvent the issue of high dimensionality.
In contrast, topology-aware classifiers, such as CNN~\cite{li2018hierarchical}, GNN~\cite{song2018eeg}, and attention-based model~\cite{jia2020sst}, take the spatial relations of \ac{EEG} channels into account and learn \ac{EEG} representations by aggregating features from different channels.
These topology-aware classifiers can utilize not only manually designed spectral features but also information dynamically extracted from raw features in the temporal domain.
 
To verify the effectiveness of brain signals, we apply two prevalent models from different groups: topology-invariant classifier \ac{DT} and topology-aware model \ac{SST}.
\ac{DT} is widely used in machine learning tasks since it can automatically choose and combine the \ac{EEG} features considering their correlation to the predicted value.
In our experiment, only spectral features are fed into \ac{DT} classifier to avoid the issue of high dimensionality. 
\ac{SST} applies attention mechanisms to adaptively capture discriminative patterns in spectral and temporal information, which achieves state-of-art performance for \ac{EEG}-based prediction tasks.
It can automatically capture \ac{EEG} features in adjacent or symmetrical channels, which may reveal crucial related information~\cite{schmidt2001frontal}.

As for the modeling of content and context information, we compare the performance of \ac{DT}, \ac{MLP}, and \ac{SVM} in our dataset.
Among them, \ac{DT} achieves the best performance, which is in consistent with previous work~\cite{mao2017understanding}.
Due to the page limits, we only report the experimental results of \ac{DT}.
Also, we perform a grid search using a trade-off parameter $\lambda$~(19 values from .05 to .95) to combine the estimation scores of models based on content/context information and brain signals. 
Then we report the performance of the combination model with different settings for the trade-off parameter $\lambda$. 

\subsection{Definitions}
To avoid ambiguity, we use $M^f$ to denote the model $M$~($=DT, SST$) using features $f$~($=cn, cx, sst$). 
\textit{cn}, \textit{cx},  and \textit{bs} indicate content features, context features, and brain signal features, respectively. 
$+$ denotes the combination of different models using trade-off parameters $\lambda$. 
For example, $DT^{cn,cx}+SST^{bs}$ denotes the combination model of \ac{DT} using features of content and context features and \ac{SST} using brain signals.

\subsection{Training Strategies and Evaluation}
Given content/context features, brain signal's features, or their combination, the task is to estimate the usefulness level of non-click search results.
We simplify the task as a binary classification problem by only considering usefulness ratings of 1 and 4.
The reasons are two-fold: (1)~Ratings of 1~(``not useful at all'') and 4~(``very useful'') are boundary usefulness judgments, and thus they contain less noise than ratings of 2 and 3. （2)~Ratings of 1 and 4 make up of 71.2\% search results in total.

To verify the performance in different application scenarios, we perform two training strategies in our experiments: \textit{user-independent} and \textit{task-independent}.  
The user-independent strategy learns a supervised model using the remaining participants' data when validating data from each participant. 
The task-independent strategy partitions the tasks into ten folds then uses the rest folds for training when validating each fold.

As for evaluation metrics, we follow the same principle as in \cite{mao2017understanding} and also use \ac{AUC} for our task to deal with the issue of imbalanced classes and report the  standard deviation of \ac{AUC} among different folds to verify the stability.


\section{Results and Analysis}

\begin{table}[]
\caption{The performance of the usefulness estimation with different information sources. $M^f$ denotes model $M$ using features $f$. \textit{cn}, \textit{cx},  and \textit{bs} indicate content, context, and brain signals, respectively. $+$ denotes grid search combination. $*$/$**$ indicate the difference of performance with $DT^{cn,cx} + SST^{bs}$ is significant with p-value $\textless$ 0.05/0.01.}
\label{tab:result}
\begin{tabular}{lp{1.15cm}p{1.15cm}p{1.2cm}p{1.2cm}}
\hlinew{0.8pt} 
\multirow{2}{*}{Model} & \multicolumn{2}{l}{\textbf{user-independent}} & \multicolumn{2}{l}{\textbf{task-independent}} \\
& \makecell[c]{AUC} & \makecell[c]{STD}  & \makecell[c]{AUC}  & \makecell[c]{STD}              \\ \hline
$DT^{cn}$ & \makecell[l]{\ \  $0.619^{**}$} & \makecell[l]{\ \  0.040} & \makecell[l]{\ \ \  $0.593^{**}$} & \makecell[l]{\ \ \  0.080}             \\
$DT^{cx}$ & \makecell[l]{\ \   $0.664^{**}$} & \makecell[l]{\ \  0.047} & \makecell[l]{\ \ \  $0.585^{**}$} & \makecell[l]{\ \ \  0.049}  \\  
$DT^{bs}$ & \makecell[l]{\ \   $0.585^{**}$} & \makecell[l]{\ \  0.047} & \makecell[l]{\ \ \  0.642} & \makecell[l]{\ \ \  0.033}  \\
$SST^{bs}$ & \makecell[l]{\ \   $0.654^{**}$} & \makecell[l]{\ \  0.043} & \makecell[l]{\ \ \  0.655} & \makecell[l]{\ \ \  0.037}             \\
$DT^{cn,cx}$ & \makecell[l]{\ \   $0.672^{**}$} & \makecell[l]{\ \  0.049} & \makecell[l]{\ \ \  $0.614^{*}$} & \makecell[l]{\ \ \  0.067} \\  
$DT^{cn,cx} + DT^{bs}$ & \makecell[l]{\ \   0.687**} & \makecell[l]{\ \  0.049} & \makecell[l]{\ \ \  0.683} & \makecell[l]{\ \ \  0.049}   \\     
$DT^{cn,cx} + SST^{bs}$ & \makecell[l]{\ \  \textbf{0.718}} & \makecell[l]{\ \  0.040} & \makecell[l]{\ \ \  \textbf{0.687}} & \makecell[l]{\ \ \  0.050}   \\
\hlinew{0.8pt}        
\end{tabular}
\end{table}

In this section, we report experimental results to answer \textbf{RQ2} and \textbf{RQ3}.
Primarily, we elaborate the overall performance of usefulness estimation with different information sources to demonstrate the effectiveness of brain signals aimed at addressing \textbf{RQ2}.
We then provide extensive analysis to study the contribution of different information sources and experimental settings~(e.g., task difficulty and length of time interval) to answer \textbf{RQ3}.

\subsection{Overall Performance}
\label{Overall Performance}
Table~\ref{tab:result} shows the overall performance of the usefulness estimation of different models on the basis of various sources~(i.e., content, context, brain signals, and their combination).
From Table~\ref{tab:result}, we have the following observations:

(1)~For both training strategies, models utilizing all features achieve significantly better performance than models that do not consider brain signals.
The best performance is achieved by $DT^{cn,cx} + SST^{bs}$ in which we use \ac{SST}  for brain signals modeling and combine it with $DT^{cn,cx}$.
This observation demonstrates that brain signals complement conventional information, including content and context features, and are beneficial for usefulness estimation. 

(2)~Compared between models utilized \ac{EEG}-based features, \ac{SST} performs better than \ac{DT}, especially in the  user-independent strategy, which indicates that considering the topological structure and utilizing attention mechanisms in brain signals modeling is beneficial.

(3)~For \ac{EEG} models, the performance in user-independent strategy is worse than that in task-independent strategy, especially for \ac{DT}.
Moreover, the standard variance in user-independent strategy is also larger than that in task-independent.
The reasons can be two folds.
On the one hand, previous work has demonstrated the phenomenon of ``BCI illiteracy''~\cite{vidaurre2010towards}, suggesting that about $15-30\%$ of persons perform poorly in \ac{BCI} systems. 
On the other hand, the brain signals across users might exist neurological differences~\cite{zheng2016personalizing}.
Our experimental results suggest that these impacts~(i.e., BCI illiteracy and neurological differences across users) might be smaller in the deep network of \ac{SST} than that in traditional classifiers \ac{DT}.


(4)~As for models excluding brain signals, they perform worse in the training strategy of task-independent than user-independent.
The reason is that some of the content and context features~(e.g., BM25 rank) are associated with the task. 
Thus it can not correlate well with unseen tasks.
However, models using brain signals do not perform worse in task-independent strategy than in user-independent strategy.
The reason can be that these models can directly capture user's psychological feedback, which is less affected by the tasks than models only applied conventional features.

\paragraph{Answer to \textbf{RQ2}}
According to the experimental results, the usefulness of non-click search results can be estimated with the help of content/context information and brain signals.
We find that additionally incorporating brain signals can improve performance significantly. 
When modeling the brain signals, models considering the topological structure and utilizing attention mechanisms~(e.g., \ac{SST}) achieve better performance than traditional topology-invariant classifiers. 
As for the training strategies, conventional features perform better in user-independent strategy than in task-independent strategy while brain signals are on the contrary.
Despite this, when applying \ac{SST} for brain signals modeling, the performance is similar across training strategies. 
This finding suggests that we should consider the modeling of various information sources for different situations, e.g., unseen users and unseen tasks.

\subsection{In-depth Analysis}

\begin{figure}
    \hspace*{\fill}%
    \subcaptionbox{User-independent.\label{fig:grid_LOPO}}
    {\includegraphics[width=.45\linewidth]{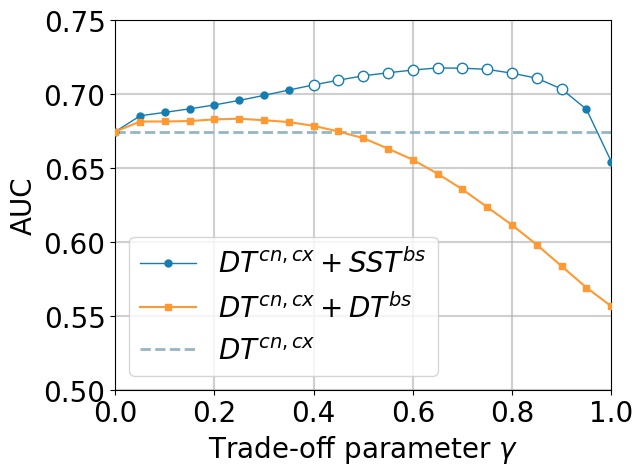}}
    \hfill\hfill\hfill\hfill%
    \subcaptionbox{Task-independent.
    \label{fig:grid_CVOQ}}
    {\includegraphics[width=.45\linewidth]{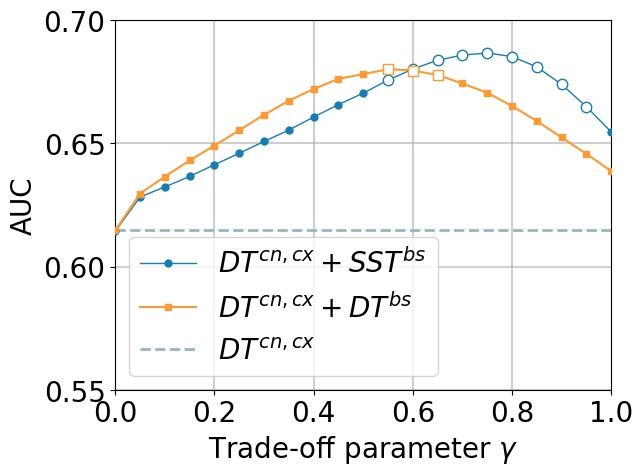}}%
    \hspace*{\fill}%
    \caption{The performance of the usefulness estimation with different trade-off parameter $\gamma$. When $\gamma$ = 0, the performance coincides with $DT^{cn,cx}$. When $\gamma$ = 1, the performance coincides with $DT^{bs}$ or $SST^{bs}$. The hollow dot denotes the performance is significantly better than $DT^{cn,cx}$.}
   	\label{fig:grid}
\end{figure}

\begin{figure}
    \hspace*{\fill}%
    \subcaptionbox{User-independent.\label{fig:difficulity_lopo}}
    {\includegraphics[width=.45\linewidth]{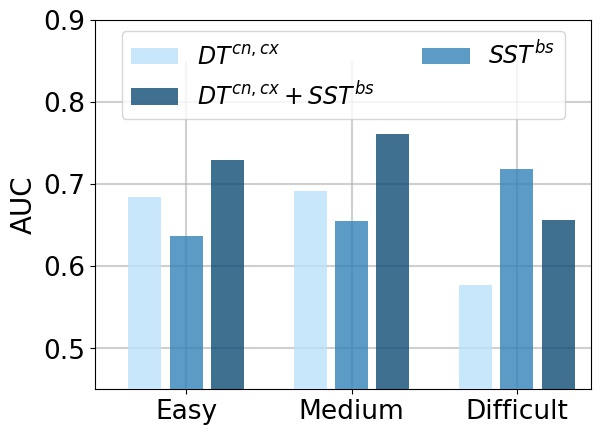}}
    \hfill\hfill\hfill\hfill%
    \subcaptionbox{Task-independent.
    \label{fig:difficulity_cvoq}}
    {\includegraphics[width=.45\linewidth]{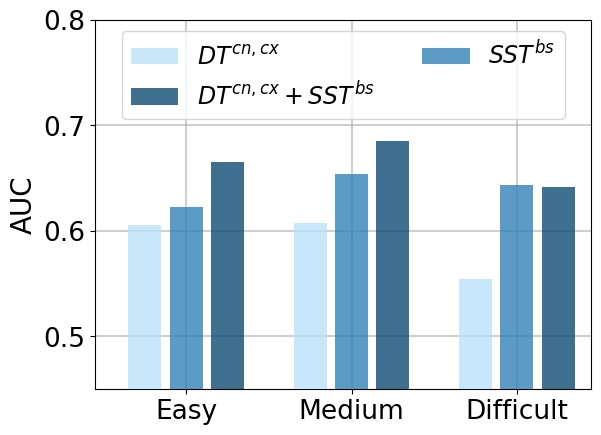}}%
    \hspace*{\fill}%
    \caption{The performance of the usefulness estimation with various task difficulty. 
    }
    \label{fig:difficulity}
\end{figure}

\subsubsection{Analysis of information sources}
By using a trade-off parameter $\gamma$ to combine the scores estimated by information sources of brain signals and content/context features, we aim to test (1)~for which settings of $\gamma$ the combination model performs better than $DT^{cn,cx}$ significantly and (2)~whether the combination model is sensitive to the $\gamma$ or not.
In Figure~\ref{fig:grid}, we show the performance of models~($DT^{cn,cx}+DT^{bs}$ and $DT^{cn,cx}+SST^{bs}$) using all features with different trade-off parameter $\gamma$.
Recall that if $\gamma$ = 0, models degrades to the $DT^{cn,cx}$, which uses content and context features only. 
While if $\gamma$ = 1, models coincide with that of $DT^{bs}$ or $SST^{bs}$ using brain signals only. 
Since \ac{SST} performs better for brain signals modeling, we mainly discuss the combination model of $DT^{cn,cx}+SST^{bs}$ and have two main observations. 

On the one hand, as $\gamma$ increases, $DT^{cn,cx}+SST^{bs}$ monotonically increases to the best performance at first and then gradually decreases after an optimal value of $\gamma$.
This finding demonstrates that incorporating conventional information and brain signals together is better than considering one facet only.

On the other hand, $DT^{cn,cx}+SST^{bs}$ is significantly better than $DT^{cn,cx}$ for $0.4 \le \gamma \le 0.85$~(user-independent) and $0.55 \le \gamma \le 0.9$~(task-independent).
However, changing $\gamma$ in $0.15 \le \gamma \le 0.85$ shows no significant differences~(for both training strategies).
These suggest that the combination model is not sensitive to this parameter.


\subsubsection{Analysis of task difficulty}
Task difficulty, which refers to a user’s assessment of the effort required to complete a search task~\cite{arguello2014predicting}, is a vital factor for search engine optimization.
The data of perceived task difficulty is collected and classified into three groups: easy~(very easy and easy), medium~(neither easy nor difficult), and difficult~(difficult and very difficult). 
On this basis, we calculate the performance of usefulness estimation, as shown in Figure~\ref{fig:difficulity}. 

The performance of $DT^{cn,cx}$ is worse in the difficult tasks than that in the easy tasks in both training strategies.
Especially in the training strategy of user-independent, repeated measures ANOVA shows that there exists a significant difference among task difficulty levels~(F[20,2]=4.24,p<0.05).
The post-hoc Bonferroni test further reveals that the model performs worse in difficult tasks than that in easy and medium tasks~(p<0.05), respectively.
This observation suggests that models using conventional features have trouble in usefulness estimation in difficult tasks.

In contrast, the performance of models based on brain signals does not decrease along with the increase of task difficulty.
Especially in user-independent strategy, although not significant, the performance in difficult tasks is even slightly better than that in easy and medium search tasks.
This finding indicates that models using brain signals are effective and robust in difficult tasks.

\begin{figure}
    \hspace*{\fill}%
    \subcaptionbox{User-independent.\label{fig:time_windowLOPO}}
    {\includegraphics[width=.45\linewidth]{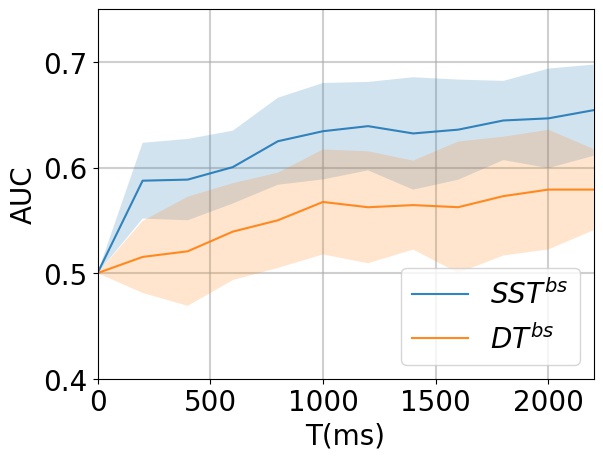}}
    \hfill\hfill\hfill\hfill%
    \subcaptionbox{Task-independent.
    \label{fig:time_windowCVOQ}}
    {\includegraphics[width=.45\linewidth]{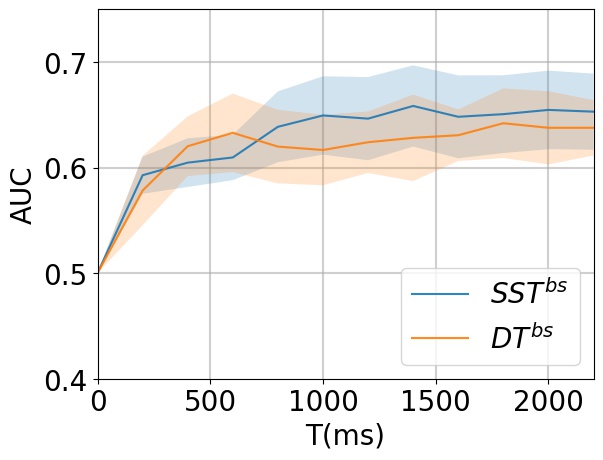}}%
    \hspace*{\fill}%
    \caption{The performance of the usefulness estimation using brain signals with different time intervals of $[0,T]$.}
    \label{fig:time_window}
\end{figure}
\subsubsection{Analysis of time intervals}
\label{6.2}
Since brain activities are time-sensitive, we further explore the influences of brain signals extracted from the different lengths of time intervals on the model performance. 
Figure~\ref{fig:time_window} shows the experimental results for both \ac{EEG}-based models.
The time intervals are $[0,T]$, where $T$ ranges from 0ms to 2500ms.
Recall that when $T=0$, the performance coincides with an untrained model.
For $SST^{bs}$ and $DT^{bs}$, we can see that the increase of performance slows down as the time interval prolongs.   
There exists no significant difference in terms of the model performance after 800ms in both training strategies.
This finding approximates an average time duration for the usefulness judgment in the result-level.
It is consistent with existing work that suggests that our brain needs around 800ms to judge the relevance of a visually presented stimulus~\cite{allegretti2015relevance}.



\paragraph{Answer to \textbf{RQ3}}
The above analyses probe into the contribution of information sources and 
experimental settings in usefulness estimation performance.
As for the combination of information sources, we demonstrate the effectiveness and stability of incorporating brain signals into the model construction.
As for experimental settings, we find that models with conventional features do not perform well in difficult tasks, but models using brain signals remain effective when search tasks become difficult.
This finding indicates that brain signals can complement conventional features and might be more beneficial for understanding user behavior in difficult search tasks.   
Finally, we suggest that the length of time interval should be at least 800ms for predicting usefulness judgments.
It implies a recommended time duration for designing a real-time feedback collecting system for search scenarios.

\section{conclusion}
This paper introduces the challenges of understanding non-click results in "Zero-click" Web search scenarios and proposes a solution using neuroimaging techniques.
Especially, we design a lab-based user study to collect brain signals of Web search users.
Based on the collected \ac{EEG} data, we find that the \ac{EEG} spectrum is significantly correlated with usefulness judgment in various brain regions~(e.g., left temporal, occipital) and spectral bands.
These findings indicate that usefulness judgments are associated with several cognitive functions which are also related to relevance perception and the satisfaction decoding in human brain. 

Inspired by these findings, we conduct extensive experiments on usefulness estimation for non-click results based on brain signals and conventional features, including content and context factors. 
Main findings include: (1)~brain signals are effective features for usefulness estimation and more robust than conventional features in different training strategies, i.e., user-independent and task-independent; (2)~the combination of models using brain signals and conventional features achieve the best performance and is not sensitive to the trade-off parameter; (3)~the performance of models only using conventional features degrades in difficult tasks while models based on brain signals do not; (4)~to better estimate usefulness, the length of the time interval for usefulness feedback collection should be more than 800ms. 

Several limitations guide interesting directions for future work: (1)~ In this paper, we perform lab-based settings in our user study.
	Analyzing brain signals in real-life search scenarios with portable \ac{EEG} devices is an interesting future work.
	(2)~ We demonstrate that brain signals are valuable signals for the usefulness estimation and can be collected almost in real-time. 
	Hence, it might be able to design scene-adaptive methods based on brain signals for real-time proactive \ac{IR} systems.

\end{CJK}
\bibliographystyle{ACM-Reference-Format}
\normalem
\balance
\bibliography{references}

\end{document}